\documentclass[aps,pra,twocolumn,amsfonts,amssymb,amsmath,showpacs,
floatfix,nofootinbib,groupedaddress,superscriptaddress,citesort]{revtex4}
\usepackage{mathrsfs}
\usepackage{amsfonts}
\usepackage{amstext}
\usepackage{amsmath}
\usepackage{amssymb}
\usepackage{bm}% bold math
\usepackage{bbm}
\usepackage[dvips]{graphicx}
\def\qed{\leavevmode\unskip\penalty9999 \hbox{}\nobreak\hfill
     \quad\hbox{\leavevmode  \hbox to.77778em{%
              \hfil\vrule   \vbox to.675em%
               {\hrule width.6em\vfil\hrule}\vrule\hfil}}
     \par\vskip3pt}

\begin{document}

%\preprint{APS/123-QED}
\title{Entangled bases with fixed Schmidt number\\}% Force line breaks with \\

\author{Yu Guo}
%\email{guoyu3@aliyun.com}
\affiliation{School of Mathematics and Computer Science, Shanxi Datong University, Datong, Shanxi 037009, China}%

\author{Shuanping Du}
%\email{dushuanping@aliyun.com}
\affiliation{School of Mathematical Sciences, Xiamen University, Xiamen, Fujian 361000,  China}

\author{Xiulan Li}
\affiliation{School of Mathematics and Computer Science, Shanxi Datong University, Datong, Shanxi 037009, China}%

\author{Shengjun Wu}
%\email{sjwu@nju.edu.cn}
\affiliation{Kuang Yaming Honors School, Nanjing University, Nanjing, Jiangsu 210093, China}

\begin{abstract}

An entangled basis with fixed Schmidt number $k$ (EBk) is a set of orthonormal
basis states with the same Schmidt number $k$ in a product Hilbert space $\mathbb{C}^d\otimes\mathbb{C}^{d'}$.
It is a generalization of both the product basis and the maximally entangled basis.
We show here that, for any $k\leq\min\{d,d'\}$, EBk
exists in $\mathbb{C}^d\otimes\mathbb{C}^{d'}$ for any $d$ and $d'$.
% ii) SEBk (EBk with the same Schmidt coefficients) exists in $\mathbb{C}^d\otimes\mathbb{C}^{d'}$ when $dd'$ is a multiple of $k$;
%iii) SEB2 exists for any $d$ and $d'$ and SEB3 exists in $\mathbb{C}^4\otimes\mathbb{C}^{d'}$ for any $d'\geq4$;
%iv) SEBk exists in $\mathbb{R}^d\otimes\mathbb{R}^{d'}$ only when $k=2^s$, $s\geq0$, and $dd'$ is a multiple of $k$.
Consequently,
general methods of constructing SEBk (EBk with the same Schmidt coefficients) and EBk (but not SEBk) are proposed.
Moreover, we extend the concept of EBk to multipartite case and find out that the multipartite
EBk can be constructed similarly.
%exists in $\mathbb{C}^d\otimes\mathbb{C}^{d'}$
%when $dd'$ is a multiple of $k$.

%\begin{description}
%\item[PACS numbers]03.67.Mn, 03.65.Db, 03.65.Ud.
%\end{description}
\end{abstract}

\maketitle
%\end{CJK*}
%\tableofcontents

\section{Introduction}

Entanglement is a fundamental feature
of quantum physics,
and it is also proved to be a central resource in quantum information
and quantum computation \cite{Horodecki2009,Guhne}.
Consequently, characterizing entanglement is one of the fundamental problem in this field.
One indispensable approach is to analyze the bases of the state space, such as
the unextendible product basis (UPB) \cite{Bennett1999}, the
unextendible maximally entangled basis (UMEB)~\cite{Bravyi},
and the unextendible entangled basis with Schmidt number $k$ (UEBk)~\cite{Guowu2014}, etc.

Bipartite basis (complete or incomplete) in $\mathbb{C}^d\otimes\mathbb{C}^{d'}$
is a fundamental problem in both quantum physics and mathematics.
The unextendible product
basis is a set of
orthogonal product vectors whose complementary space
does not contain product states, which
can be used for
constructing bound entangled states \cite{Pittenger}.
A UMEB is a set of orthonormal maximally entangled pure states in a two-qudit system consisting of fewer than
$d^2$ members which have no other maximally entangled vectors orthogonal to all of them.
It is shown that there is no UMEB in the two-qubit system, a six-member UMEB exists in $\mathbb{C}^3\otimes \mathbb{C}^3$ and
a 12-member UMEB exists in $\mathbb{C}^4\otimes \mathbb{C}^4$~\cite{Bravyi}.
Later, B. Chen and S.-M. Fei proved in Ref. \cite{Chen} that there exists a set of $d^2$-member UMEB in $\mathbb{C}^d\otimes \mathbb{C}^{d'}$ ($\frac{d'}{2}<d<d'$) and
questioned the existence of UMEBs in the case of $d'\geq 2d$. Recently,
in Ref. \cite{Limaosheng}, the authors proved that there might be two sets of UMEBs in any bipartite system, and
an explicit construction of UMEBs is put forward.
Some properties of UMEBs are given in Ref.~\cite{Wangyanling}.

One of the most crucial quantity associated with bipartite pure state is the \emph{Schmidt number},
which can be used to
characterize and quantify the degree of bipartite entanglement for
pure states directly~\cite{Donald,Sperling}.
In Ref. \cite{Guowu2014}, we introduce the unextendible entangled basis with Schmidt number $k$ ($2\leq k<\min\{d,d'\}$) (UEBk)
and a general way of constructing such a basis with arbitrary $d$ and $d'$ is proposed.
Consequently, it is shown that there are at least $k-r$ (here $r=d$ mod $k$, or $r=d'$ mod $k$) sets of UEBk when $d$ or $d'$ is not the multiple of $k$, while
there are at least $2(k-1)$ sets of UEBk when both $d$ and $d'$ are the multiples of $k$.
UEBk can be considered as a generalization of both UPB and UMEB from different directions.
In Ref.~\cite{Baumgartner06,Baumgartner07}, it is shown that there exists MEB (maximally entangled basis) in $\mathbb{C}^d\otimes\mathbb{C}^{d}$
for any $d$.
We remark that the product basis (PB) always exists in $\mathbb{C}^d\otimes\mathbb{C}^{d'}$ for arbitrary $d$ and $d'$ and
it is a set of pure states whose Schmidt numbers are 1. Furthermore, the MEB is set of pure states whose
Schmidt numbers are d.
Then, a related question is arisen:
is there an entangled basis with any fixed Schmidt number in $\mathbb{C}^d\otimes\mathbb{C}^{d'}$?
In this paper, we show that such bases exist in both $\mathbb{C}^d\otimes\mathbb{C}^{d'}$
and $\mathbb{R}^d\otimes\mathbb{R}^{d'}$ for any $d$ and
$d'$, and we provide methods to construct them.
We also show that our methods can be directly extended to the multipartite case.
Such bases could be useful when one study projective measurements onto basis states with a fixed Schmidt number.

The material in this paper is arranged as follows. In
Sec. II, we introduce the concepts of EBk, SEBk and MEB.
In addition the relation between EBk and the rank-$k$ Hilbert-Schmidt basis of the associated matrix space is illustrated.
Sec. III contains methods of
constructing EBk whenever $dd'$ is a multiple of $k$. In Sec. IV, we
discuss the case when $dd'$ is not a multiple of $k$ by analyzing the Hilbert-Schmidt basis with fixed rank $k$ in the space
of coefficients matrices.
Both EBk and SEBk are discussed.
%The case of $\mathbb{R}^d\otimes\mathbb{R}^{d'}$ is also considered.
The multipartite case is discussed in Sec. V.
Finally,
we conclude in Sec. VI.

\section{Definition and preliminary}

Recall that, the Schmidt number of a pure state $|\psi\rangle\in\mathbb{C}^d\otimes\mathbb{C}^{d'}$, denote by $S_r(|\psi\rangle)$, is defined
as the length of the Schmidt decomposition \cite{Schmidt}:
if
$|\psi\rangle=\sum_{k=0}^{m-1}\lambda_{k}|e_k\rangle|e_k'\rangle$ is its
Schmidt decomposition, then
$S_r(|\psi\rangle)=m$. It is clear that
$S_r(|\psi\rangle)={\rm rank}(\rho_1)={\rm rank}(\rho_2)$,
where $\rho_i$ denotes the reduced state of the $i-$th part.
A state $|\psi\rangle\in\mathbb{C}^d\otimes\mathbb{C}^{d'}$ is called a maximally entangled state if
$S_r(|\psi\rangle)=d$ and $\lambda_1=\lambda_2=\cdots=\lambda_d$.
Hereafter, we always assume that $d\leq d'$ for simplicity.

{\it Definition 1.} An orthonormal basis $\{|\phi_i\rangle\}$ in $\mathbb{C}^d\otimes\mathbb{C}^{d'}$ is called an
entangled basis with Schmidt number $k$ (EBk) if $S_r(|\phi_i\rangle)=k\geq 2$ for any $i$.
Particularly, it is called a special entangled basis with Schmidt number $k$ (SEBk) if it is an EBk with all the Schmidt coefficients of $|\psi_i\rangle$s equal to $\frac{1}{\sqrt{k}}$,
and it is called a maximally entangled basis (MEB) if $|\phi_i\rangle$ is maximally entangled for any $i$.

PB always exists in $\mathbb{C}^d\otimes\mathbb{C}^{d'}$ for any $d$ and $d'$.
It is clear that SEBk reduces to PB (MEB) when $k=1$ ($k=d$).
That is, SEBk is a generalization of both PB and MEB.
MEB is equal to SEBd while an EBd is not a MEB necessarily.
By definition, it is clear that $\{|\phi_i\rangle\}$ is an EBk if and only if
$\{U_a\otimes U_b|\phi_i\rangle\}$ is an EBk, where $U_{a/b}$ is any unitary operator on $\mathbb{C}^{d/d'}$.

We assume that $\{|\psi_i\rangle\}_{i=0}^{dd'-1}$ is set of pure states in $\mathbb{C}^d\otimes\mathbb{C}^{d'}$.
Let
\begin{eqnarray*}
|\psi_i\rangle=\sum_{k,l}a^{(i)}_{kl}|k\rangle|l'\rangle,
\end{eqnarray*}
where $\{|k\rangle\}$ and
$|l'\rangle$ are the standard computational bases of the first $\mathbb{C}^d$ and the second $\mathbb{C}^{d'}$,
respectively. We write
\begin{eqnarray*}
A_i=[a_{kl}^{(i)}],
\end{eqnarray*}
then $A_i$ is a $d$ by $d'$ matrix,
$S_r(|\psi_i\rangle)={\rm rank}(A_i)$
and $\langle\psi_i|\psi_j\rangle=\text{Tr}(A_i^\dag A_j)$.
Let $\mathcal{M}_{d\times d'}$ be the space of all $d$ by $d'$ complex matrices.
Then $\mathcal{M}_{d\times d'}$ is a Hilbert space with the inner product defined by $\langle A|B\rangle={\rm Tr}(A^{\dag}B)$ for any $A$, $B\in\mathcal{M}_{d\times d'}$.
It turns out that $\{A_i:\text{rank}(A_i)=k\}$ is a Hilbert-Schmidt basis of the space of $\mathcal{M}_{d\times d'}$ if and only if $\{|\psi_i\rangle\}_{i=0}^{dd'-1}$
is an EBk in $\mathbb{C}^d\otimes\mathbb{C}^{d'}$.
For simplicity, we call $\{A_i:\text{rank}(A_i)=k, \ {\rm Tr}(A_i^{\dag}A_j)=\delta_{ij}\}$ a rank-$k$ basis in the following.
That is, there is a one-to-one relation between the EBk $\{|\psi_i\rangle\}$ and the rank-$k$ basis $\{A_i\}$:
\begin{eqnarray}
|\psi_i\rangle\leftrightarrow A_i, \
\{|\psi_i\rangle\}\leftrightarrow \{A_i\}.
\label{relation}
\end{eqnarray}
Therefore, the EBk problem is equivalent to the rank-$k$ basis of the associated matrix space.
The important role of  Eq.~(\ref{relation}) will be seen in Sec. IV for the case when $dd'$ is not a multiple of $k$, although the case when $dd'$ is a multiple of $k$ in the next section can be worked out without using Eq.~(\ref{relation}).

\section{$dd'$ is a multiple of $k$}

This section is divided into three cases. We firstly discuss the case $k=d=d'$, then
consider the case $k=d< d'$, and study the case $k<d\leq d'$ at last.
The methods of constructing EBk can be induced from the structure of the MEB.

\subsection{$k=d=d'$}

For the two-qubits case, the Bell basis states,
$\frac{1}{\sqrt{2}}(|0\rangle|0'\rangle+|1\rangle|1'\rangle)$, $\frac{1}{\sqrt{2}}(|0\rangle|0'\rangle-|1\rangle|1'\rangle)$,
$\frac{1}{\sqrt{2}}(|0\rangle|1'\rangle+|1\rangle|0'\rangle)$, and $\frac{1}{\sqrt{2}}(|0\rangle|1'\rangle-|1\rangle|0'\rangle)$
form a MEB.
In general, for the case of $d\otimes d$, $d\geq2$,
let
\begin{eqnarray*}
|\Omega_{0,0}\rangle=\frac{1}{\sqrt{d}}\sum\limits_{i=0}^{d-1}|i\rangle|i'\rangle
\end{eqnarray*}
and
\begin{eqnarray}
\check{W}_{m,n}|i\rangle=\xi^{m(i-n)}|i-n\rangle,
\label{weyloperation}
\end{eqnarray}
where $\check{W}_{m,n}$s are the Weyl operators, $\xi=e^{\frac{2\pi{\rm i}}{d}}$
and $|i-n\rangle\equiv|i-n+d\rangle$ (here, the roman letter `i' denotes the imaginary unit, $i-n+d$ means $i-n+d$ mod $d$).
Then the actions of the Weyl operators produce a MEB~\cite{Baumgartner06,Baumgartner07}.
\begin{eqnarray}
|\Omega_{m,n}\rangle=(\check{W}_{m,n}\otimes \mathbbm{1})|\Omega_{0,0}\rangle,
\end{eqnarray}
where $0\leq m,n\leq d-1$.
In fact, Eq.~(\ref{weyloperation}) can be simplified as
\begin{eqnarray}
\hat{W}_{m,n}|i\rangle=\xi^{mi}|i-n\rangle.
\label{weyloperation2}
\end{eqnarray}
Then
\begin{eqnarray}
|\hat{\Omega}_{m,n}\rangle=(\hat{W}_{m,n}\otimes \mathbbm{1})|\Omega_{0,0}\rangle
\label{meb}
\end{eqnarray}
with $0\leq m,n\leq d-1$
also form a MEB in $\mathbb{C}^d\otimes \mathbb{C}^d$.
Symmetrically,
both  $(\mathbbm{1}\otimes\check{W}_{m,n})|\Omega_{0,0}\rangle$ and
 $(\mathbbm{1}\otimes\hat{W}_{m,n})|\Omega_{0,0}\rangle$ induce a MEB as well.

\subsection{$k=d< d'$}

We begin with the simple case of $2\otimes 3$.
It is clear that
\begin{eqnarray*}
|\phi_1\rangle&=&\frac{1}{\sqrt{2}}(|0\rangle|0'\rangle+|1\rangle|1'\rangle),\\
|\phi_2\rangle&=&\frac{1}{\sqrt{2}}(|0\rangle|0'\rangle-|1\rangle|1'\rangle),\\
|\phi_3\rangle&=&\frac{1}{\sqrt{2}}(|0\rangle|1'\rangle+|1\rangle|2'\rangle),\\
|\phi_4\rangle&=&\frac{1}{\sqrt{2}}(|0\rangle|1'\rangle-|1\rangle|2'\rangle),\\
|\phi_5\rangle&=&\frac{1}{\sqrt{2}}(|0\rangle|2'\rangle+|1\rangle|0'\rangle),\\
%\end{eqnarray*}
%\begin{eqnarray*}
|\phi_6\rangle&=&\frac{1}{\sqrt{2}}(|0\rangle|2'\rangle-|1\rangle|0'\rangle)
\end{eqnarray*}
is a MEB in $\mathbb{C}^2\otimes \mathbb{C}^3$.
In general, in a $d\otimes d'$ ($d< d'$) system,
let
\begin{eqnarray*}
|\Omega_{0,0}\rangle=\frac{1}{\sqrt{d}}\sum\limits_{i=0}^{d-1}|i\rangle|i'\rangle
\end{eqnarray*}
and
\begin{eqnarray}
\tilde{W}_{m,n}|i'\rangle=\xi^{mi}|(i-n)'\rangle,
\label{weyloperation3}
\end{eqnarray}
where $\xi=e^{\frac{2\pi{\rm i}}{d}}$
and $|i-n\rangle\equiv|i-n+d'\rangle$ (here $i-n+d'$ means $i-n+d'$ mod $d'$).
Then
\begin{eqnarray}
|\tilde{\Omega}_{m,n}\rangle=( \mathbbm{1} \otimes \tilde{W}_{m,n})|\Omega_{0,0}\rangle
\label{meb2}
\end{eqnarray}
with $0\leq m \leq d-1$ and $0\leq n\leq d'-1$ induce a MEB in $\mathbb{C}^d\otimes \mathbb{C}^{d'}$.

That is, SEBd always exists in $\mathbb{C}^d\otimes \mathbb{C}^{d'}$ with $d\leq d'$.
We can also show that there is EBd but not SEBd in $\mathbb{C}^d\otimes \mathbb{C}^{d'}$.
Let
\begin{eqnarray*}
O_d=\frac{1}{d}\left(\begin{array}{ccccc}
2-d&2&2&\cdots&2 \\
2&2-d&2&\cdots&2 \\
2&2&2-d&\cdots&2 \\
\vdots&\vdots&\vdots&\ddots&\vdots\\
2&2&2&\cdots&2-d\end{array}\right),
\end{eqnarray*}
then $O_d$ is a $d$ by $d$ orthogonal matrix and $U_d={\rm i}O_d$ is a unitary matrix.
Replacing the coefficients in Eq.~(\ref{meb},\ref{meb2}) by the entries in the columns (or rows) of the $U_d$ or $O_d$ respectively,
one can get the existence of EBd in any $d\otimes d'$ system.

In fact, any $d$ by $d$ isometric matrix  $X=[x_{kl}]$ without zero entries can induce an EBd since
we can replace the coefficients in Eq.~(\ref{meb},\ref{meb2}) by the entries in the columns of any $d$ by $d$ isometric matrix without zero entries.
That is, we replace the coefficient $\xi^{mi}$ of $\xi^{mi}|i\rangle|(n-i)'\rangle$ by $x_{n+1,m+1}$.
Here, an $m$ by $n$ matrix $A$ is an isometric matrix if $A^\dag A=I_n$, $I_n$ is the $n$ by $n$ identity matrix.
Since there are infinitely many isometric matrices, we can construct infinitely many EBds.
Also note that, there are many ways of constructing MEB since we can replacing the coefficient $\xi^{mi}$ by any other ones that
guarantee the orthogonality.

\subsection{$k<d\leq d'$}

We now consider the EBk in a $d\otimes d'$ system with $dd'$ a multiple of $k$, and $k<d\leq d'$.
For clarity, we give an example of EB3 in $\mathbb{C}^4\otimes \mathbb{C}^{6}$.
Let $\xi=e^{\frac{2\pi{\rm i}}{3}}$, it is obvious that the following states constitute
an EB3 in a $4\otimes 6$ system.
\begin{eqnarray*}
|\phi_0\rangle&=&\frac{1}{\sqrt{3}}(|0\rangle|0'\rangle+|1\rangle|1'\rangle+|2\rangle|2'\rangle),\\
|\phi_1\rangle&=&\frac{1}{\sqrt{3}}(|0\rangle|0'\rangle+\xi|1\rangle|1'\rangle+\xi^2|2\rangle|2'\rangle),\\
|\phi_2\rangle&=&\frac{1}{\sqrt{3}}(|0\rangle|0'\rangle+\xi^2|1\rangle|1'\rangle+\xi^4|2\rangle|2'\rangle),\\
|\phi_3\rangle&=&\frac{1}{\sqrt{3}}(|3\rangle|3'\rangle+|0\rangle|1'\rangle+|1\rangle|2'\rangle),\\
|\phi_4\rangle&=&\frac{1}{\sqrt{3}}(|3\rangle|3'\rangle+\xi|0\rangle|1'\rangle+\xi^2|1\rangle|2'\rangle),\\
|\phi_5\rangle&=&\frac{1}{\sqrt{3}}(|3\rangle|3'\rangle+\xi^2|0\rangle|1'\rangle+\xi^4|1\rangle|2'\rangle),\\
|\phi_6\rangle&=&\frac{1}{\sqrt{3}}(|2\rangle|3'\rangle+|3\rangle|4'\rangle+|0\rangle|2'\rangle),
\end{eqnarray*}
\begin{eqnarray*}
|\phi_7\rangle&=&\frac{1}{\sqrt{3}}(|2\rangle|3'\rangle+\xi|3\rangle|4'\rangle+\xi^2|0\rangle|2'\rangle),\\
|\phi_8\rangle&=&\frac{1}{\sqrt{3}}(|2\rangle|3'\rangle+\xi^2|3\rangle|4'\rangle+\xi^4|0\rangle|2'\rangle),\\
|\phi_9\rangle&=&\frac{1}{\sqrt{3}}(|1\rangle|3'\rangle+|2\rangle|4'\rangle+|3\rangle|5'\rangle),\\
|\phi_{10}\rangle&=&\frac{1}{\sqrt{3}}(|1\rangle|3'\rangle+\xi|2\rangle|4'\rangle+\xi^2|3\rangle|5'\rangle),\\
|\phi_{11}\rangle&=&\frac{1}{\sqrt{3}}(|1\rangle|3'\rangle+\xi^2|2\rangle|4'\rangle+\xi^4|3\rangle|5'\rangle),\\
|\phi_{12}\rangle&=&\frac{1}{\sqrt{3}}(|0\rangle|3'\rangle+|1\rangle|4'\rangle+|2\rangle|5'\rangle),\\
|\phi_{13}\rangle&=&\frac{1}{\sqrt{3}}(|0\rangle|3'\rangle+\xi|1\rangle|4'\rangle+\xi^2|2\rangle|5'\rangle),\\
|\phi_{14}\rangle&=&\frac{1}{\sqrt{3}}(|0\rangle|3'\rangle+\xi^2|1\rangle|4'\rangle+\xi^4|2\rangle|5'\rangle),\\
|\phi_{15}\rangle&=&\frac{1}{\sqrt{3}}(|3\rangle|0'\rangle+|0\rangle|4'\rangle+|1\rangle|5'\rangle),\\
|\phi_{16}\rangle&=&\frac{1}{\sqrt{3}}(|3\rangle|0'\rangle+\xi|0\rangle|4'\rangle+\xi^2|1\rangle|5'\rangle),\\
|\phi_{17}\rangle&=&\frac{1}{\sqrt{3}}(|3\rangle|0'\rangle+\xi^2|0\rangle|4'\rangle+\xi^4|1\rangle|5'\rangle),\\
|\phi_{18}\rangle&=&\frac{1}{\sqrt{3}}(|2\rangle|0'\rangle+|3\rangle|1'\rangle+|0\rangle|5'\rangle),\\
|\phi_{19}\rangle&=&\frac{1}{\sqrt{3}}(|2\rangle|0'\rangle+\xi|3\rangle|1'\rangle+\xi^2|0\rangle|5'\rangle),\\
|\phi_{20}\rangle&=&\frac{1}{\sqrt{3}}(|2\rangle|0'\rangle+\xi^2|3\rangle|1'\rangle+\xi^4|0\rangle|5'\rangle),\\
|\phi_{21}\rangle&=&\frac{1}{\sqrt{3}}(|1\rangle|0'\rangle+|2\rangle|1'\rangle+|3\rangle|2'\rangle),\\
|\phi_{22}\rangle&=&\frac{1}{\sqrt{3}}(|1\rangle|0'\rangle+\xi|2\rangle|1'\rangle+\xi^2|3\rangle|2'\rangle),\\
|\phi_{23}\rangle&=&\frac{1}{\sqrt{3}}(|1\rangle|0'\rangle+\xi^2|2\rangle|1'\rangle+\xi^4|3\rangle|2'\rangle).
\end{eqnarray*}
With the same spirit in mind, in general, we let
\begin{eqnarray*}
|\gamma_i\rangle=\left\{\begin{array}{ll}
|i\rangle|i'\rangle,&0\leq i< d,\\
|r\rangle|(t\oplus r)'\rangle,& i=td+r, 0\leq r<d,
\end{array}\right.
\end{eqnarray*}
where $t\oplus r$ means $t+r$ mod $d'$.
If $dd'=sk$, $2\leq k< d$, then
\begin{eqnarray}
|\breve{\Omega}_{m,n}\rangle=\frac{1}{\sqrt{k}}\sum\limits_{l=0}^{k-1}\xi^{ml}|\gamma_{nk+l}\rangle,
\label{ebk}
\end{eqnarray}
constitute an EBk, where $\xi=e^{\frac{2\pi{\rm i}}{k}}$, $0\leq m\leq k-1$, $0\leq n\leq s-1$.

The EBk above is also a SEBk. In order to construct an EBk that is not a SEBk, one can replace the coefficients in Eq.~(\ref{ebk}) by
the entries of any $k$ by $k$ isometric matrix $X=[x_{ij}]$ without zero entries,
namely, we can replace the coefficient $\xi^{ml}$ in Eq.~(\ref{ebk}) by $x_{l+1,m+1}$.
From the discussion in Subsec. A and Subsec. B we know that we can construct
infinitely many such EBks that are not SEBks, and we also know that there are other methods of constructing SEBks.

\section{$dd'$ is not a multiple of $k$}

In this section, we assume that $dd'$ is not a multiple of $k$.
We discuss an EBk that is not a SEBk firstly and then
deal with SEBks.
%We will discuss  by analyzing the structure of the coefficient matrices of the basis states.

We begin with the case $k=2$.
If $dd'$ is not a multiple of $2$, then the Hilbert space $\mathcal{M}_{d\times d'}$
is a direct sum of three subspaces which are equivalent to
$\mathcal{M}_{2p\times 3}$, $\mathcal{M}_{d\times 2q}$ and $\mathcal{M}_{3\times 3}$ respectively, with $p,q\geq 1$.
For example, the space of $\mathcal{M}_{3\times 5}$ is a direct sum of
$\mathcal{M}_{3\times 2}$ and $\mathcal{M}_{3\times 3}$:
\begin{eqnarray*}
\left(\begin{array}{ccccc}
*&*&*&*&* \\
*&*&*&*&* \\
*&*&*&*&* \end{array}\right)=
\left(\begin{array}{ccccc}
*&*&0&0&0 \\
*&*&0&0&0 \\
*&*&0&0&0 \end{array}\right)\oplus
\left(\begin{array}{ccccc}
0&0&*&*&* \\
0&0&*&*&* \\
0&0&*&*&* \end{array}\right);
\end{eqnarray*}
the space of $\mathcal{M}_{5\times 5}$ is a direct sum of
$\mathcal{M}_{5\times 2}$, $\mathcal{M}_{2\times 3}$ and $\mathcal{M}_{3\times 3}$:
\begin{eqnarray*}
\left(\begin{array}{ccccc}
*&*&*&*&* \\
*&*&*&*&* \\
*&*&*&*&* \\
*&*&*&*&* \\
*&*&*&*&*\end{array}\right)=
\left(\begin{array}{ccccc}
*&*&0&0&0 \\
*&*&0&0&0 \\
*&*&0&0&0 \\
*&*&0&0&0 \\
*&*&0&0&0\end{array}\right)~~~~~~~~~~~~\\
\oplus
\left(\begin{array}{ccccc}
0&0&0&0&0 \\
0&0&0&0&0 \\
0&0&0&0&0 \\
0&0&*&*&* \\
0&0&*&*&*\end{array}\right)\oplus
\left(\begin{array}{ccccc}
0&0&*&*&* \\
0&0&*&*&* \\
0&0&*&*&* \\
0&0&0&0&0 \\
0&0&0&0&0\end{array}\right).
\end{eqnarray*}
Since rank-2 basis always exists in $\mathcal{M}_{2p\times s}$ and $\mathcal{M}_{t\times 2q}$,
we only need to discuss $\mathcal{M}_{3\times 3}$.
Observe that
\begin{eqnarray*}
\left(\begin{array}{ccc}
*&*&* \\
*&*&* \\
*&*&* \end{array}\right)=
\left(\begin{array}{ccc}
0&*&* \\
*&0&* \\
*&0&* \end{array}\right)\oplus
\left(\begin{array}{ccc}
*&0&0 \\
0&*&0 \\
0&*&0 \end{array}\right),
\end{eqnarray*}
we thus only need to check
\begin{eqnarray*}
\mathcal{L}_{2+1}=\left\{\left(\begin{array}{cc}
*&0 \\
0&* \\
0&*\end{array}\right)\right\}
\end{eqnarray*}
have rank-2 bases since $\left(\begin{array}{ccc}
0&*&* \\
*&0&* \\
*&0&* \end{array}\right)$ have rank-2 bases (it is indeed a $2\otimes 3$ case).
Similarly, for $k=3$,
the Hilbert space $\mathcal{M}_{d\times d'}$
is a direct sum of three subspaces which are equivalent to
$\mathcal{M}_{d\times 3p}$, $\mathcal{M}_{3q\times (3+t)}$ and $\mathcal{M}_{(3+s)\times (3+t)}$ respectively,
where $p,q\geq 1$, $1\leq s,t\leq 2$.
We thus only need to show
\begin{eqnarray*}
\mathcal{L}_{3+1}=\left\{\left(\begin{array}{ccc}
*&0&0 \\
0&*&0 \\
0&0&*\\
0&0&*\end{array}\right)\right\} \text{ and } \mathcal{L}_{3+2}=\left\{\left(\begin{array}{ccc}
*&0&0 \\
0&*&0 \\
0&0&*\\
0&0&*\\
0&0&*
\end{array}\right)\right\}
\end{eqnarray*}
have rank-3 bases.
In general, for any $k$, if $d=sk+r$ and $d'=s'k+r'$, then
the Hilbert space $\mathcal{M}_{d\times d'}$
is a direct sum of three subspaces which are equivalent to
$\mathcal{M}_{d\times (s'-1)k}$, $\mathcal{M}_{(s-1)k\times (k+r')}$ and $\mathcal{M}_{(k+r)\times (k+r')}$ respectively.
We only need to consider rank-$k$ bases in the following spaces
(assume that there exists EBl $(l<k)$)
\begin{eqnarray*}
\mathcal{L}_{k+1}=\left\{\left(\begin{array}{ccccc}
x_1&0&\cdots&0&0 \\
0&x_2&\cdots&0&0 \\
\vdots&\vdots&\ddots&\vdots&\vdots\\
0&0&\cdots&0&x_k\\
0&0&\cdots&0&x_{k+1}\end{array}\right)\right\},
\end{eqnarray*}
\begin{eqnarray*}
\mathcal{L}_{k+2}=\left\{\left(\begin{array}{ccccc}
x_1&0&\cdots&0&0 \\
0&x_2&\cdots&0&0 \\
\vdots&\vdots&\ddots&\vdots&\vdots\\
0&0&\cdots&0&x_k\\
0&0&\cdots&0&x_{k+1}\\
0&0&\cdots&0&x_{k+2}\end{array}\right)\right\},
\end{eqnarray*}
\begin{eqnarray*}
\cdots%~~~~~~~~~~~~~~~~~~
\end{eqnarray*}
\begin{eqnarray*}
\mathcal{L}_{2k-1}=\left\{\left(\begin{array}{ccccc}
x_1&0&\cdots&0&0 \\
0&x_2&\cdots&0&0 \\
\vdots&\vdots&\ddots&\vdots&\vdots\\
0&0&\cdots&0&x_k\\
0&0&\cdots&0&x_{k+1}\\
0&0&\cdots&0&x_{k+2}\\
\vdots&\vdots&\ddots&\vdots&\vdots\\
0&0&\cdots&0&x_{2k-1}
\end{array}\right)\right\}.
\end{eqnarray*}
In fact,
for the space of $\mathcal{L}_{k+s}$, $1\leq s\leq k-1$,
we take
\begin{eqnarray}
x_{p}^{(i)}=
\frac{1}{\sqrt{k+s}}\xi^{(p-1)(i-1)},\ \xi=e^{\frac{2\pi{\rm i}}{k+s}},
\label{ebk2}
\end{eqnarray}
$1\leq p,i\leq k+s$.
That is, $x_p^{(i)}$s are the coefficients of the $i$-th element of
MEB in $(k+s)\otimes(k+s)$ system as in Eq.~(\ref{meb}), $i=1$, 2, $\dots$, $k+s$.
Let
\begin{eqnarray}
A_{k+s}^{(i)}=\left(\begin{array}{ccccc}
x_1^{(i)}&0&\cdots&0&0 \\
0&x_2^{(i)}&\cdots&0&0 \\
\vdots&\vdots&\ddots&\vdots&\vdots\\
0&0&\cdots&0&x_k^{(i)}\\
0&0&\cdots&0&x_{k+1}^{(i)}\\
0&0&\cdots&0&x_{k+2}^{(i)}\\
\vdots&\vdots&\ddots&\vdots&\vdots\\
0&0&\cdots&0&x_{k+s}^{(i)}
\end{array}\right),
\label{p-A}
\end{eqnarray}
then $\{A_{k+s}^{(i)}\}$ is a rank-$k$ basis of $\mathcal{L}_{k+s}$.
We thus prove that EBk exits in any bipartite system.

The coefficients in Eq.~(\ref{ebk2}) can be replaced by the entries
in the columns of any $k+s$ by $k+s$ isometric matrix without zero entries respectively.
That is, there are infinitely many EBks in any $d\otimes d'$ system.
Of course there also exist other types of EBk (but not SEBk).
For example,
\begin{eqnarray}
|\psi_1\rangle&=&\frac{1}{2}|0\rangle|0'\rangle+\frac{\sqrt{3}}{2}|1\rangle|1'\rangle,\nonumber \\
|\psi_2\rangle&=&\frac{\sqrt{3}}{2}|0\rangle|0'\rangle-\frac{1}{2}|1\rangle|1'\rangle,\nonumber \\
%\end{eqnarray*}
%\begin{eqnarray*}
|\psi_3\rangle&=&\frac{1}{\sqrt{2}}|0\rangle|1'\rangle+\frac{1}{\sqrt{2}}|1\rangle|0'\rangle,\nonumber \\
|\psi_4\rangle&=&-\frac{1}{\sqrt{2}}|0\rangle|1'\rangle+\frac{1}{\sqrt{2}}|1\rangle|0'\rangle,\nonumber \\
|\psi_5\rangle&=&\frac{1}{\sqrt{2}}|1\rangle|2'\rangle+\frac{1}{\sqrt{2}}|2\rangle|0'\rangle,\nonumber \\
|\psi_6\rangle&=&\frac{1}{\sqrt{2}}|1\rangle|2'\rangle-\frac{1}{\sqrt{2}}|2\rangle|0'\rangle,\nonumber \\
|\psi_7\rangle&=&\frac{1}{\sqrt{2}}|0\rangle|2'\rangle+\frac{1}{\sqrt{2}}|2\rangle|1'\rangle,\nonumber \\
|\psi_8\rangle&=&\frac{1}{\sqrt{3}}|0\rangle|2'\rangle-\frac{1}{\sqrt{3}}|2\rangle|1'\rangle+\frac{1}{\sqrt{3}}|2\rangle|2'\rangle,\nonumber \\
|\psi_9\rangle&=&\frac{1}{\sqrt{6}}|0\rangle|2'\rangle-\frac{1}{\sqrt{6}}|2\rangle|1'\rangle-\frac{\sqrt{2}}{\sqrt{3}}|2\rangle|2'\rangle
\label{eb2}
\end{eqnarray}
constitute an EB2 in $\mathbb{C}^3\otimes\mathbb{C}^3$, but it is not a SEB2.
We thus obtain the following result.

{\it Theorem 1.} For any $d$, $d'$ and $1\leq k\leq d$, there exist infinitely many
EBks in $\mathbb{C}^d\otimes\mathbb{C}^{d'}$.

Now we discuss SEBk. We begin with $k=2$ and $d=d'=3$.
It is straightforward to show that, for any rank-2 basis of $\mathcal{L}_{2+1}$, if it corresponds
to three elements of a SEB2, then it
must admit the following form
\begin{eqnarray*}
c\left(\begin{array}{cc}
e^{{\rm i}x_1}&0\\
0&ae^{{\rm i}x_2} \\
0&be^{{\rm i}x_2} \end{array}\right),
c\left(\begin{array}{cc}
e^{{\rm i}y_1}&0 \\
0&ae^{{\rm i}y_2} \\
0&be^{{\rm i}y_2} \end{array}\right),
%\end{eqnarray*}
%and
%\begin{eqnarray*}
c\left(\begin{array}{cc}
e^{{\rm i}z_1}&0 \\
0&ae^{{\rm i}z_2} \\
0&be^{{\rm i}z_2} \end{array}\right),
\end{eqnarray*}
where $a>0$, $b>0$, $a^2+b^2=1$, $c=\frac{\sqrt{2}}{2}$,
$x_i$, $y_i$, $z_i$ ($i=1, 2$) are real numbers that should satisfy the following equalities
\begin{eqnarray*}
\left\{\begin{array}{c}
x_1-y_1=x_2-y_2\pm \pi,\\
x_1-z_1=x_2-z_2\pm \pi,\\
y_1-z_1=y_2-z_2\pm \pi.\end{array}\right.
\end{eqnarray*}
However, the above three equalities contradict each other. Therefore,
there is no SEB2 in $3\otimes 3$ systems according to our Scenario (maybe SEB2 can be constructed
by other approaches).
The space of $\mathcal{M}_{4\times 4}$ can be decomposed as
\begin{eqnarray*}
\left(\begin{array}{cccc}
*&*&*&* \\
*&*&*&* \\
*&*&*&* \\
*&*&*&* \end{array}\right)=
\left(\begin{array}{cccc}
0&*&*&* \\
*&0&*&* \\
*&*&0&* \\
*&*&*&0 \end{array}\right)\oplus
\left(\begin{array}{cccc}
*&0&0&0 \\
0&*&0&0 \\
0&0&*&0 \\
0&0&0&* \end{array}\right).
\end{eqnarray*}
\if We only need to consider
$\left(\begin{array}{cccc}
*&0&0&0 \\
0&*&0&0 \\
0&0&*&0 \\
0&0&0&* \end{array}\right)$
since there exist rank-3 bases in the space
$\left\{\left(\begin{array}{cccc}
0&*&*&* \\
*&0&*&* \\
*&*&0&* \\
*&*&*&0 \end{array}\right)\right\}$.
\fi
It is obvious that
${\rm diag}($0,1,1,1),
${\rm diag}$(1,0,-1,1),
${\rm diag}$(1,1,0,-1)
and
${\rm diag}$(1,-1,1,0)
form a rank-3 basis
of $\left\{{\rm diag}(*,*,*,*)\right\}$.
Thus, SEB3 exists in any $4\otimes d'$ systems or systems that can be divided into
$3\otimes p$, $4\otimes q$, $r\otimes 4$ and $s\otimes 3$ systems with $q, r\geq 4$ and $p,s\geq 3$.
%We conjecture that there exist SEBk for any $k$.
We now can conclude the following.

{\it Theorem 2.} If $dd'$ is a multiple of $k$, then there exist infinitely many
SEBks in $\mathbb{C}^d\otimes\mathbb{C}^{d'}$.

{\it Theorem 3.} If $dd'$ is not a multiple of $k$, $d=sk+r$ and $d'=s'k+r'$, then there exists SEBk in $\mathbb{C}^d\otimes\mathbb{C}^{d'}$
if there exists a $(k+\check{r})\times (k+\check{r})$ isometric matrix $X=[x_{ij}]$, $\check{r}=\min\{r,r'\}$,
such that each column
\begin{eqnarray*}
(x_{1j},x_{2j},\cdots,x_{k+s,j})^t\quad
\end{eqnarray*}
(here, $A^t$ denotes the transpose of $A$) satisfies:

 1) either there are only $k$ nonzero entries and the modulus of them are $\frac{1}{\sqrt{k}}$ or

 2) $|x_{ij}|=\frac{1}{\sqrt{k}}$ when $1\leq i\leq k-1$ and $\sum_{i\geq k}|x_{ij}|^2=\frac{1}{k}$.

That is, the SEBk problem is reduced to the construction of the special isometric matrices.
However,
whether or not there exists $k+s$ by $k+s$ isometric matrix $X=[x_{ij}]$ that satisfying the condition 1) or 2) is
unknown when $k\geq 3$.
With the increase of dimensions $d$ and $d'$ and $k$, the verification of the existence of SEBk becomes harder and harder.

\section{Multipartite case}

In this section, we consider the multipartite case.
We firstly extend the concept of EBk to multipartite systems.
In a product Hilbert space $\mathbb{C}^{d_1}\otimes\mathbb{C}^{d_2}\otimes\cdots\otimes\mathbb{C}^{d_N}$ with $N\geq3$, only specific pure states admit a Schmidt
decomposition form~\cite{Peres1995,Thapliyal}. Now we discuss whether a basis can be constructed from such specific states for the multipartite case.
Hereafter, we always assume with no loss of generality that $d_1\leq d_2\leq\cdots\leq d_N$.

{\it Definition 2.} An orthonormal basis $\{|\psi_i\rangle\}$ in $\mathbb{C}^{d_1}\otimes\mathbb{C}^{d_2}\otimes\cdots\otimes\mathbb{C}^{d_N}$ is
a $N$-partite EBk ($1\leq k\leq d_1$) if
\begin{eqnarray}
|\psi_i\rangle=\sum_{j=0}^{k-1}\lambda_j^{(i)}|e_{j}^{(1)}\rangle|e_{j}^{(2)}\rangle\cdots|e_{j}^{(N)}\rangle
\end{eqnarray}
for any $i$, $0\leq i\leq d_1d_2\cdots d_N-1$, where $\{|e_{j}^{(l)}\rangle\}$ is an orthonormal
set of $\mathbb{C}^{d_l}$, $\lambda_j^{(i)} >0$, $\sum_j(\lambda_j^{(i)})^2=1$, $1\leq l\leq N$, $N\geq 3$.
Particularly, it is a $N$-partite SEBk if it is a $N$-partite EBk with all the coefficients $\lambda_j^{(i)}$s equal to $\frac{1}{\sqrt{k}}$.

We begin with the tripartite case. Let $\{|\psi_i\rangle\}$ be an EBk of $\mathbb{C}^{d_1}\otimes\mathbb{C}^{d_2}$, and be written in the Schmidt decomposition form as $|\psi_i\rangle=\sum_{j=0}^{k-1}\lambda_j^{(i)}|e_j\rangle|e_{j}'\rangle$, of which the $l$-th product term $|e_l\rangle|e_{l}'\rangle$ is denoted by $|\psi_i^{l}\rangle$ for convenience.
We define
\begin{eqnarray}
|\psi_{i,j}^{(2+1)}\rangle:=\sum_{l=0}^{k-1}\lambda_l^{(i)}|\psi_i^{l}\rangle|(j+l)_3\rangle,
\label{3partite}
\end{eqnarray}
where $j+l$ means $j+l$ mod $d_3$, $\{|j_3\rangle\}$ is an orthonormal basis of $\mathbb{C}^{d_3}$, $0\leq j\leq d_3-1$.
Then $\{|\psi_{i,j}^{2+1}\rangle\}$ with $0\leq i\leq d_1d_2-1$ and $0\leq j\leq d_3-1$ is a tripartite EBk in $\mathbb{C}^{d_1}\otimes\mathbb{C}^{d_2}\otimes\mathbb{C}^{d_3}$.

Generally, we denote by $|\psi^{l}\rangle=|e_{l}^{(1)}\rangle|e_{l}^{(2)}\rangle\cdots|e_{l}^{(m)}\rangle$ for convenience provided that $|\psi\rangle=\sum_{j=0}^{k-1}\lambda_j|e_{j}^{(1)}\rangle|e_{j}^{(2)}\rangle\cdots|e_{j}^{(m)}\rangle$ with $\{|e_{j}^{(s)}\rangle\}$ is an orthonormal
set of $\mathbb{C}^{d_s}$, $1\leq s\leq m$.
If $\{|\psi_i\rangle\}$ is an $m$-partite EBk of $\mathbb{C}^{d_1}\otimes\mathbb{C}^{d_2}\otimes\cdots\otimes\mathbb{C}^{d_m}$,
then
\begin{eqnarray}
|\psi_{i,j}^{(m+1)}\rangle:=\sum_{l=0}^{k-1}\lambda_l^{(i)}|\psi_i^{l}\rangle|(j+l)_{m+1}\rangle
\label{mpartite}
\end{eqnarray}
with $0\leq j\leq d_{m+1}-1$ form an $m+1$-partite EBk of $\mathbb{C}^{d_1}\otimes\mathbb{C}^{d_2}\otimes\cdots\otimes\mathbb{C}^{d_{m+1}}$,
where $\{|j_{m+1}\rangle\}$ is an orthonormal basis of $\mathbb{C}^{d_{m+1}}$, $j+l$ means $j+l$ mod $d_{m+1}$, $0\leq i\leq d_1d_2\cdots d_m-1$.
That is, $(m+1)$-partite EBk can be obtained from $m$-partite EBk for any $m\geq2$.
Together with Theorems 1-3, we thus get the following.

{\it Proposition 1.} For any $1\leq k\leq d_1$, there exist infinitely many
$N$-partite EBks in $\mathbb{C}^{d_1}\otimes\mathbb{C}^{d_2}\otimes\cdots\otimes\mathbb{C}^{d_N}$.

{\it Proposition 2.} If $d_1d_2\cdots d_N$ is a multiple of $k$, then there exist infinitely many
$N$-partite SEBks in $\mathbb{C}^{d_1}\otimes\mathbb{C}^{d_2}\otimes\cdots\otimes\mathbb{C}^{d_N}$.

{\it Proposition 3.} If $d_1d_2\cdots d_N$ is not a multiple of $k$, then there exists $N$-partite SEBk in $\mathbb{C}^{d_1}\otimes\mathbb{C}^{d_2}\otimes\cdots\otimes\mathbb{C}^{d_N}$
if the condition of Theorem 3 holds with $\check{r}=\min\{r_i:1\leq i\leq N\}$, $d_i=s_ik+r_i$.

In other words, the key point of multipartite case is in fact the same as the bipartite one.
In addition, from our approach, the $N$-partite EBk can be constructed in many ways.
For example, if $d_2d_3$ is a multiple of $k$, one can either construct an EBk (or SEBk)
of $\mathbb{C}^{d_2}\otimes\mathbb{C}^{d_3}$ or construct an EBk (or SEBk)
of $\mathbb{C}^{d_p}\otimes\mathbb{C}^{d_q}$ ( $(p,q)\neq (2,3)$), then a $N$-partite EBk (or SEBk) is obtained by Eq.~(\ref{mpartite}).
%We now complete our analyzing for any systems with any dimension.

The basis vector $|\psi_i\rangle$ of any $N$-partite EBk $\{|\psi_i\rangle\}$
is in fact a GHZ-like state, namely,
it is local unitarily equivalent to the form
\begin{eqnarray*}
|{\rm GHZ}_k^{(N)}\rangle=\sum\limits_{j=0}^{k-1}\lambda_j|j_1\rangle|j_2\rangle\cdots|j_{N}\rangle,
\end{eqnarray*}
where $\{|j_l\rangle\}$ is an orthonormal
set of $\mathbb{C}^{d_l}$, $\lambda_j>0$, $\sum_j\lambda_j^2=1$, $1\leq l\leq N$.

At last, we give two examples for illustration.
For the case $\mathbb{C}^2\otimes\mathbb{C}^2\otimes\mathbb{C}^2$, it is easy to check that
the eight 3-qubits GHZ states deduced from the four Bell states via relation Eq.~(\ref{3partite}) form a 3-partite SEB2 (or MEB),
and a general 3-partite EB2 (that is not a SEB2) can be obtained through Eq.~(\ref{3partite}) from any EB2 in $\mathbb{C}^2\otimes\mathbb{C}^2$.
Namely, if $\{|\psi_i\rangle=a_i^{(0)}|\psi_i^0\rangle+a_i^{(1)}|\psi_i^1\rangle: 0\leq i\leq 3,\ (a_i^{(0)})^2+(a_i^{(1)})^2=1\}$ is an EB2 of two-qubits system, then
\begin{eqnarray*}
|\psi_{0,0}^{(2+1)}\rangle&=&a_0^{(0)}|\psi_0^0\rangle|0\rangle+a_0^{(1)}|\psi_0^1\rangle|1\rangle,\\
|\psi_{0,1}^{(2+1)}\rangle&=&a_0^{(0)}|\psi_0^0\rangle|1\rangle+a_0^{(1)}|\psi_0^1\rangle|0\rangle,\\
|\psi_{1,0}^{(2+1)}\rangle&=&a_1^{(0)}|\psi_1^0\rangle|0\rangle+a_1^{(1)}|\psi_1^1\rangle|1\rangle,\\
|\psi_{1,1}^{(2+1)}\rangle&=&a_1^{(0)}|\psi_1^0\rangle|1\rangle+a_1^{(1)}|\psi_1^1\rangle|0\rangle,
\end{eqnarray*}
\begin{eqnarray*}
|\psi_{2,0}^{(2+1)}\rangle&=&a_2^{(0)}|\psi_2^0\rangle|0\rangle+a_2^{(1)}|\psi_2^1\rangle|1\rangle,\\
|\psi_{2,1}^{(2+1)}\rangle&=&a_2^{(0)}|\psi_2^0\rangle|1\rangle+a_2^{(1)}|\psi_2^1\rangle|0\rangle,\\
|\psi_{3,0}^{(2+1)}\rangle&=&a_3^{(0)}|\psi_3^0\rangle|0\rangle+a_3^{(1)}|\psi_3^1\rangle|1\rangle,\\
|\psi_{3,1}^{(2+1)}\rangle&=&a_3^{(0)}|\psi_3^0\rangle|1\rangle+a_3^{(1)}|\psi_3^1\rangle|0\rangle
\end{eqnarray*}
are a 3-partite EB2 in $\mathbb{C}^2\otimes\mathbb{C}^2\otimes\mathbb{C}^2$ via relation Eq.~(\ref{3partite}).

For the case $\mathbb{C}^3\otimes\mathbb{C}^3\otimes\mathbb{C}^3$,
\begin{eqnarray*}
|\psi_{0,0}^{(2+1)}\rangle&=&\frac{1}{2}|0\rangle|0\rangle|0\rangle+\frac{\sqrt{3}}{2}|1\rangle|1\rangle|1\rangle,\\
|\psi_{0,1}^{(2+1)}\rangle&=&\frac{1}{2}|0\rangle|0\rangle|1\rangle+\frac{\sqrt{3}}{2}|1\rangle|1\rangle|2\rangle,\\
|\psi_{0,2}^{(2+1)}\rangle&=&\frac{1}{2}|0\rangle|0\rangle|2\rangle+\frac{\sqrt{3}}{2}|1\rangle|1\rangle|0\rangle,\\
|\psi_{1,0}^{(2+1)}\rangle&=&\frac{\sqrt{3}}{2}|0\rangle|0\rangle|0\rangle-\frac{1}{2}|1\rangle|1\rangle|1\rangle,\\
|\psi_{1,1}^{(2+1)}\rangle&=&\frac{\sqrt{3}}{2}|0\rangle|0\rangle|1\rangle-\frac{1}{2}|1\rangle|1\rangle|2\rangle,\\
|\psi_{1,2}^{(2+1)}\rangle&=&\frac{\sqrt{3}}{2}|0\rangle|0\rangle|2\rangle-\frac{1}{2}|1\rangle|1\rangle|0\rangle,\\
|\psi_{2,0}^{(2+1)}\rangle&=&\frac{1}{\sqrt{2}}|0\rangle|1\rangle|0\rangle+\frac{1}{\sqrt{2}}|1\rangle|0\rangle|1\rangle,\\
|\psi_{2,1}^{(2+1)}\rangle&=&\frac{1}{\sqrt{2}}|0\rangle|1\rangle|1\rangle+\frac{1}{\sqrt{2}}|1\rangle|0\rangle|2\rangle,\\
|\psi_{2,2}^{(2+1)}\rangle&=&\frac{1}{\sqrt{2}}|0\rangle|1\rangle|2\rangle+\frac{1}{\sqrt{2}}|1\rangle|0\rangle|0\rangle,\\
%\end{eqnarray*}
%\begin{eqnarray*}
|\psi_{3,0}^{(2+1)}\rangle&=&-\frac{1}{\sqrt{2}}|0\rangle|1\rangle|0\rangle+\frac{1}{\sqrt{2}}|1\rangle|0\rangle|1\rangle,\\
|\psi_{3,1}^{(2+1)}\rangle&=&-\frac{1}{\sqrt{2}}|0\rangle|1\rangle|1\rangle+\frac{1}{\sqrt{2}}|1\rangle|0\rangle|2\rangle,\\
|\psi_{3,2}^{(2+1)}\rangle&=&-\frac{1}{\sqrt{2}}|0\rangle|1\rangle|2\rangle+\frac{1}{\sqrt{2}}|1\rangle|0\rangle|0\rangle,\\
|\psi_{4,0}^{(2+1)}\rangle&=&\frac{1}{\sqrt{2}}|1\rangle|2\rangle|0\rangle+\frac{1}{\sqrt{2}}|2\rangle|0\rangle|1\rangle,\\
|\psi_{4,1}^{(2+1)}\rangle&=&\frac{1}{\sqrt{2}}|1\rangle|2\rangle|1\rangle+\frac{1}{\sqrt{2}}|2\rangle|0\rangle|2\rangle,\\
|\psi_{4,2}^{(2+1)}\rangle&=&\frac{1}{\sqrt{2}}|1\rangle|2\rangle|2\rangle+\frac{1}{\sqrt{2}}|2\rangle|0\rangle|0\rangle,\\
|\psi_{5,0}^{(2+1)}\rangle&=&\frac{1}{\sqrt{2}}|1\rangle|2\rangle|0\rangle-\frac{1}{\sqrt{2}}|2\rangle|0\rangle|1\rangle,\\
|\psi_{5,1}^{(2+1)}\rangle&=&\frac{1}{\sqrt{2}}|1\rangle|2\rangle|1\rangle-\frac{1}{\sqrt{2}}|2\rangle|0\rangle|2\rangle,\\
|\psi_{5,2}^{(2+1)}\rangle&=&\frac{1}{\sqrt{2}}|1\rangle|2\rangle|2\rangle-\frac{1}{\sqrt{2}}|2\rangle|0\rangle|0\rangle,\\
|\psi_{6,0}^{(2+1)}\rangle&=&\frac{1}{\sqrt{2}}|0\rangle|2\rangle|0\rangle+\frac{1}{\sqrt{2}}|2\rangle|1\rangle|1\rangle,\\
|\psi_{6,1}^{(2+1)}\rangle&=&\frac{1}{\sqrt{2}}|0\rangle|2\rangle|1\rangle+\frac{1}{\sqrt{2}}|2\rangle|1\rangle|2\rangle,\\
|\psi_{6,2}^{(2+1)}\rangle&=&\frac{1}{\sqrt{2}}|0\rangle|2\rangle|2\rangle+\frac{1}{\sqrt{2}}|2\rangle|1\rangle|0\rangle,
\end{eqnarray*}
\begin{eqnarray*}
|\psi_{7,0}^{(2+1)}\rangle&=&\frac{1}{\sqrt{3}}|0\rangle|2\rangle|0\rangle
-\frac{1}{\sqrt{3}}|2\rangle|1\rangle|1\rangle+\frac{1}{\sqrt{3}}|2\rangle|2\rangle|1\rangle,\\
|\psi_{7,1}^{(2+1)}\rangle&=&\frac{1}{\sqrt{3}}|0\rangle|2\rangle|1\rangle
-\frac{1}{\sqrt{3}}|2\rangle|1\rangle|2\rangle+\frac{1}{\sqrt{3}}|2\rangle|2\rangle|2\rangle,\\
|\psi_{7,2}^{(2+1)}\rangle&=&\frac{1}{\sqrt{3}}|0\rangle|2\rangle|2\rangle
-\frac{1}{\sqrt{3}}|2\rangle|1\rangle|0\rangle+\frac{1}{\sqrt{3}}|2\rangle|2\rangle|0\rangle,\\
|\psi_{8,0}^{(2+1)}\rangle&=&\frac{1}{\sqrt{6}}|0\rangle|2\rangle|0\rangle
-\frac{1}{\sqrt{6}}|2\rangle|1\rangle|1\rangle-\frac{\sqrt{2}}{\sqrt{3}}|2\rangle|2\rangle|1\rangle,\\
|\psi_{8,1}^{(2+1)}\rangle&=&\frac{1}{\sqrt{6}}|0\rangle|2\rangle|1\rangle
-\frac{1}{\sqrt{6}}|2\rangle|1\rangle|2\rangle-\frac{\sqrt{2}}{\sqrt{3}}|2\rangle|2\rangle|2\rangle,\\
|\psi_{8,2}^{(2+1)}\rangle&=&\frac{1}{\sqrt{6}}|0\rangle|2\rangle|2\rangle
-\frac{1}{\sqrt{6}}|2\rangle|1\rangle|0\rangle-\frac{\sqrt{2}}{\sqrt{3}}|2\rangle|2\rangle|0\rangle
\end{eqnarray*}
constitute an EB2 in $\mathbb{C}^3\otimes\mathbb{C}^3\otimes\mathbb{C}^3$,
$|\psi_{i,j}^{(2+1)}\rangle$s are in fact three-qubits GHZ-like states.
It is clear that $|\psi_{i,j}^{(2+1)}\rangle$s above are derived from Eq.~(\ref{eb2}) via relation Eq.~(\ref{3partite}).
There is no SEB2 in $3\otimes 3$ system, so there is no SEB2 in $3\otimes 3\otimes 3$ system either.
An EB3 or SEB3 in $\mathbb{C}^3\otimes\mathbb{C}^3\otimes\mathbb{C}^3$ can be obtained via relation Eq.~(\ref{3partite}) from any EB3 or SEB3 in $\mathbb{C}^3\otimes\mathbb{C}^3$ respectively.

\section{Conclusions}

We have introduced the concept of EBk and showed that
EBk exists
in $\mathbb{C}^d\otimes\mathbb{C}^{d'}$ for any $d$ and $d'$.
In our discussion we have also proposed methods of constructing EBks by
analyzing the structure of the matrix space of the coefficients matrices.
We showed that the existence of EBks is equivalent to the existence of the special
isometric matrices.
Our result is a complement of the UB problem (here, UB contains UPB, UMEB and UEBk) to some extent.
By now, the basis problems (namely, MEB, EBk, PB, UPB, UMEB and UEBk) are somehow settled down:
any kind of basis (complete or incomplete) exists in any bipartite space $\mathbb{C}^d\otimes\mathbb{C}^{d'}$.
It is a little pity that the SEBk problem is left open when $dd'$ is not a multiple of $k$.%, although we believe that they exist as well.
%We stressed that the EBk of space $\mathbb{R}^d\otimes\mathbb{R}^{d'}$ is different from
%that of
%$\mathbb{C}^d\otimes\mathbb{C}^{d'}$: there is no SEBk in $\mathbb{R}^d\otimes\mathbb{R}^{d'}$ whenever $k\neq2^s$, $s\geq0$.

Furthermore, our approaches can be used in multipartite case and we thus have provided
a complete characterization of EBk in both the bipartite case and the multipartite case.
In either bipartite case or multipartite case, the basis states of EBk (or $N$-partite EBk) are just the pure states that admit the
Schmidt decomposition with the same length.
Although only some special multipartite pure states have the Schmidt decomposition,
the basis with such a special structure always exists.
Our results provide a mathematical tool for studying projective measurements onto basis states that admit a Schmidt decomposition with the same length.

It is worth mentioning here that real space is different from the complex space.
From the arguments in Sec. III-IV,  we conjecture that there is no
SEB2 in the real space $\mathbb{R}^d\otimes\mathbb{R}^{d'}$ when $dd'$ is not a multiple of $2$
(due to the fact that the real numbers with modulus 1 are only 1 and -1).
Moreover, we conjecture that: i) MEB exists only when $d=2^s$ with $s\geq 0$;
ii) SEBk exists only when $k=2^s$, $s\geq0$, and $dd'$ is a multiple of $k$.
In addition, from the orthogonal matrix $O_n$, we can conclude that EBk that is not SEBk
exists in $\mathbb{R}^d\otimes\mathbb{R}^{d'}$ for any $d$ and $d'$
(of course there are other types of EBks).
Similar results for the multipartite case can be obtained from the method in Sec. V.

\begin{acknowledgements}
We thank Valerio Scarani for useful remarks.
Guo is supported by the National Natural Science Foundation of China under Grants No. 11301312 and 11171249,
the Natural Science Foundation of Shanxi
under Grant No. 2013021001-1 and 2012011001-2, and the Research start-up fund for Doctors of Shanxi Datong University
under Grant No. 2011-B-01.
Du is supported by the National Natural Science Foundation of
China under Grant No. 11001230 and the Natural Science Foundation of Fujian under Grants No. 2013J01022 and 2014J01024.
Wu is supported by the National Natural Science Foundation of China under Grants No. 11275181 and No. 11475084.
\end{acknowledgements}

%\nocite{*}


\end{document}